%% file: main.tex
\documentclass[aps,prd,twocolumn,superscriptaddress]{revtex4-2}
\usepackage{graphicx} % Required for inserting images
\usepackage{amsmath}
\usepackage{color}
\usepackage[dvipsnames]{xcolor}
\usepackage{rotating}
\usepackage{bigints}
\usepackage{enumerate}
\usepackage{array}
\usepackage{multirow}

\newcommand{\beq}{\begin{equation}}
\newcommand{\eeq}{\end{equation}}
\newcommand{\bea}{\begin{eqnarray}}
\newcommand{\eea}{\end{eqnarray}}
\newcommand{\hmu}{\hat{\mu}}

\begin{document}

\title{Taste breaking in the minimally doubled Karsten-Wilczek action and its tree-level improvement}

\author{S. Bors\'anyi}
\affiliation{University of Wuppertal, Department of Physics, Wuppertal D-42119, Germany}

\author{S. Capitani}
\affiliation{University of Wuppertal, Department of Physics, Wuppertal D-42119, Germany}

\author{Z. Fodor}
\affiliation{Pennsylvania State University, Department of Physics, State College, PA 16801, USA}
\affiliation{University of Wuppertal, Department of Physics, Wuppertal D-42119, Germany}
\affiliation{Inst.  for Theoretical Physics, ELTE E\"otv\"os Lor\' and University, P\'azm\'any P. s\'et\'any 1/A, H-1117 Budapest, Hungary}
\affiliation{J\"ulich Supercomputing Centre, Forschungszentrum J\"ulich, D-52425 J\"ulich, Germany}

\author{D. Godzieba}
\affiliation{Pennsylvania State University, Department of Physics, State College, PA 16801, USA}

\author{P. Parotto}
\affiliation{Dipartimento di Fisica, Universit\`a di Torino and INFN Torino, Via P. Giuria 1, I-10125 Torino, Italy}

\author{R. A. Vig}
\affiliation{University of Wuppertal, Department of Physics, Wuppertal D-42119, Germany}

\author{C. H. Wong}
\affiliation{University of Wuppertal, Department of Physics, Wuppertal D-42119, Germany}

\date{\today}

\begin{abstract}
    Minimally doubled fermion actions offer a discretization for two-flavor 
    Quantum Chromodynamics without rooting, but retaining a U(1) chiral 
    symmetry at the same time. The price to pay is a breaking of the 
    hypercubic symmetry, which requires the inclusion and tuning of 
    new counterterms. 
    Similar to staggered quarks, these actions suffer from taste breaking. 
    We perform a mixed action numerical study with the Karsten-Wilczek formulation of minimally doubled fermions on 4stout staggered configurations, generated with physical quark masses, covering a broad range of lattice spacings. 
    We consider a tree-level spatial Naik improvement to mitigate discretization errors. 
    We carry out a non-perturbative tuning of the KW action
    with and without improvement, and investigate 
    the taste breaking and the approach to the continuum limit.
\end{abstract}

\maketitle 

%%%%%%%%%%%%%%%%%%%%%%%%%%%%%%%%%%%%%%%%%%%%%%%%%%%%%%%%%%%%%%%%%%%%%%%%%%%%%%%%
\section{Introduction}

Dynamical lattice simulations use a variety of discretizations of the 
Quantum Chromodynamics (QCD)
action, which differ by the features they conserve from the continuum
theory, ease of implementation and numerical cost.
Perhaps the most widespread formulation is the Wilson action where an
additional clover term is introduced to ensure that the errors come
$\mathcal{O}(a^2)$ of the lattice spacing \cite{Sheikholeslami:1985ij}.
This action has been used in several precision studies in the recent past
to calculate $n$-point functions in the QCD vacuum, e.g. hadron spectroscopy \cite{Durr:2008zz}.
On the other hand, when studying the chiral transition of finite temperature QCD,
and in particular the phase diagram of the theory, actions
which preserve a remnant chiral symmetry are preferrable, if not mandatory.
Since chiral actions are in general numerically expensive, we owe the vast 
majority of QCD thermodynamics results in the literature to their cheapest 
version, namely staggered quarks~\cite{Kogut:1974ag}. 
Because the staggered discretization suffers from four-fold doubling 
for each flavour of fermion, rooting is required on the 
quark determinant whenever the number of degenerate fermion flavours is not 
a multiple of four. The soundness of fermion rooting has been 
debated for a long 
time~\cite{Sharpe:2006re,Golterman:2006rw,Creutz:2007yg,Giordano:2019gev}.
At the very least, rooting gives a well defined procedure at vanishing
chemical potential, where the fermion determinant is real. 
This is not the case, however, when a (real) chemical potential is 
introduced, and the determinant becomes complex. 

Among the most common methods to circumvent the complex action problem
of QCD at real chemical potential are reweighting techniques~\cite{Barbour:1997ej,Fodor:2001au,Fodor:2001pe,Fodor:2004nz,deForcrand:2002pa,Alexandru:2005ix,Fodor:2007vv,Endrodi:2018zda}, which have
enjoyed promising developments in recent 
years~\cite{Giordano:2020uvk, Giordano:2020roi, Borsanyi:2021hbk,Borsanyi:2022soo}.
In the case of $N_f=2$ degenerate light flavours, the rooting at finite 
chemical potential suffers from a sign ambiguity. It was shown in 
Ref.~\cite{Borsanyi:2023tdp} that staggered rooting at finite chemical 
potential leads to unphysical effects in the thermodynamics at low 
temperatures, as a remnant of pion condensation at finite isospin chemical 
potential.

In order to avoid the rooting of the light quarks, one possibility is to employ so-called minimally 
doubled fermions. As the name suggests, minimally doubled fermions have
the minimum number of doublers allowed by the Nielsen-Ninoyima 
no-go theorem \cite{Nielsen:1981hk}, while still maintaining a remnant chiral 
symmetry with a strictly local Dirac operator. Minimally doubled fermions 
reduce the number of doublers from fifteen to one, leaving two degenerate
mass species. In the vast majority of thermodynamics studies, up and down
quarks are treated as degenerate. Hence in the $N_f=2$ theory the number of
quark states is the correct one, and no rooting procedure is necessary. In
the case of the $N_f=2+1$ theory, a rooting of the strange quark determinant
would still be necessary. However, the unphysical effects of rooting in this
case would appear at rather large strange chemical potentials, in a regime which
is beyond the reach of current reweighting methods. The remnant chiral 
symmetry is a feature that makes minimally doubled fermions particularly 
attractive for thermodynamics studies. The major drawback is that 
the resulting breaking of hypercubic symmetry requires the introduction of 
counterterms that have no counterpart in the continuum theory.
This makes the renormalization process more involved than with
other actions.
The most popular versions of minimally doubled 
fermion actions are the Karsten-Wilczek 
(KW)~\cite{Karsten:1981gd,Wilczek:1987kw} and 
Bori\c{c}i-Creutz 
(BC)~\cite{Creutz:2007af,Borici:2007kz} fermions, although other types exist, including generalizations of both \cite{Capitani:2013zta,Capitani:2013fda,Capitani:2013iha}.

The most relevant concern about minimally doubled fermions comes from the apparent $\mathcal{O}(a)$ discretization errors. The origin of this effect is the inclusion of a dimension-five
term, a 3D Laplacian operator, designed to eliminate the spatial doublers. This has
the consequence that both the self energy and the vacuum polarization diagrams
suffer from terms that are odd in $a$, similarly to the unimproved Wilson action.
It is natural to expect that an improvement program can be introduced based on the distinct
dimension-five operators of the theory \cite{Weber:2015hib}. However, even in its unimproved
form, the action has an automatic improvement feature for some applications: precisely the terms that are odd in the lattice spacing, are also odd in one of the broken symmetries (e.g.
charge conjugation or time reversal) \cite{Weber:2015oqf}. If that symmetry is restored on
the level of the observable, such terms will cancel, as long as the gauge configurations
are generated with an action that respects the discrete symmetries. This is a typical pattern in mixed action studies like the present one.

For dynamical minimally doubled fermions we know from pertrubation theory that
terms which are effectively of order $a$ appear as leading logarithms
$a \, g_0^{2n} \log^n a$ in the loop amplitudes, where $g_0$ is the bare gauge coupling,
running with the lattice spacing as $g_0^2 \sim 1 /\log a$.
As a first step towards full improvement we introduce a tree-level improved action,
which has the effect of eliminating the leading logarithms,
so that the amplitudes behave as $a \, g_0^{2n} \log^{n-1} a$
\cite{Heatlie:1990kg}. At one loop this means that the leading logarithms
$a \, g_0^2 \log a$ disappear, leaving finite $\sim a g_0^2$ terms as leading corrections.

In this work we use a Naik type (3-hop) construction to achieve tree-level improvement.
The new irrelevant term is designed to raise the dimensionality of the 3D Laplacian,
at least on a trivial gauge background. We will show in a separate publication
that this type of improvement eliminates the leading log divergences
\cite{Capitani:2025ikw}.

Though in a mixed action study odd terms in $a$ cancel on average for $C$ or $T$ symmetric observables,
the symmetry breaking terms increase the statistical errors. 
The decrease of magnitude of the radiative corrections from $a \, g_0^2 \log a$
to $a \, g_0^2$ thanks to improvement has the consequence of reducing the
statistical noise in the functional integral which is sampled
in Monte Carlo simulations, as we can see in our calculation of the
pion decay constant $f_\pi$.

In this work we present a study of the tuning procedure and the taste breaking of the Karsten-Wilczek action. We do so on a gauge background generated with the 4stout action, 
employing a tree-level Symanzik gauge action and a staggered fermionic action 
with four levels of stout smearing.
We implement both the original unimproved version of the KW action as well
as a tree level Naik improved version, discussed in Sections \ref{sec:kw} and \ref{sec:naik},
respectively.
In both cases we carry out the non-perturbative renormalization (Section \ref{sec:renorm}), and 
test the consistency of the continuum result for the pion decay constant with the staggered 
formulation (Section \ref{sec:fpi}).
Finally, we determine the taste breaking features of the action 
in both scenarios (Section \ref{sec:taste}), and discuss the corresponding computational demands in Section \ref{sec:conclusions}.

%%%%%%%%%%%%%%%%%%%%%%%%%%%%%%%%%%%%%%%%%%%%%%%%%%%%%%%%%%%%%%%%%%%%%%%%%%%%%%%%
\section{\label{sec:kw}The Karsten-Wilczek action}

The first and simplest formulation of minimally doubled fermions is due to 
Karsten~\cite{Karsten:1981gd} and Wilczek~\cite{Wilczek:1987kw}, whereby the 
single remaining doubler lies on one of the axes of the Brillouin zone. 
Like Wilson fermions this is achieved by adding a term to the na\"ive action that
is proportional to a Laplacian. Minimally doubled fermions have the extra requirement
to anticommute with $\gamma_5$, in order not to violate chiral symmetry. Thus, 
the Laplacian term cannot be added in all directions. At most a 3D Laplacian can be considered, multiplied
by the $\gamma$ matrix of the fourth direction. This construction leaves
only one doubler lying in the fourth space-time direction in the Brillouin zone. 
This design inherently breaks  the hypercubic symmetry of the lattice down to a cubic symmetry 
corresponding to the doubler-free directions. Wilczek generalized the action 
by introducing the so called Wilczek parameter $\zeta$. 

The tree-level Karsten-Wilczek action is given by:
\begin{align} \nonumber
S_F^{KW} &= S_F^N + \sum_x \bar{\psi}(x) \frac{i \zeta}{2} \gamma_\alpha \sum_{\mu \neq \alpha} \left( 2 \psi(x) - U_\mu(x) \psi(x+\hmu) \right.\\
& \left. \qquad \qquad \qquad \qquad - \, U_\mu^\dagger(x-\hmu) \psi(x-\hmu)  \right) \rm{,}
\end{align}
where $S_F^N$ is the na\"ive fermion action
\begin{align} \nonumber
S_F^N &= \sum_x \sum_{\mu} 
\bar\psi(x) \gamma_\mu \frac12 \left[U_\mu(x)
\psi(x+\hmu) \right. \\
& \qquad \left. - U^+_\mu(x-\hmu)\psi(x-\hmu)
\right] + m \sum_x \bar\psi(x)\psi(x) \, \, .
\end{align}
Here, $U_\mu(x)$ are the links in direction $\mu$ at lattice site $x$. The sum over the directions $\mu \neq \alpha$ implements a three dimensional covariant Laplacian.
This is the so-called Karsten-Wilczek term
and is analogous to a Wilson term though
with a different gamma matrix structure.
The action is minimally doubled if its coefficient, the Karsten-Wilczek parameter, obeys $|\zeta| > 1/2$.  
The different treatment of direction $\alpha$ and the three other directions
introduces an anisotropy. It is convenient to choose $\alpha$ as the temporal direction.

While the action at first sight looks more similar to Wilson fermions, due 
to an exact remnant chiral symmetry, the spectrum of the massless Dirac 
operator is on the imaginary line, similarly to na\"ive and staggered
fermions. Karsten-Wilczek fermions also perceive the global topological 
charge the same way as na\"ive and staggered fermions do~\cite{Durr:2022mnz}. 

% Similarly to staggered fermions, Karsten-Wilczek fermions suffer from tastes 
% breaking, but instead of the 4 tastes of the staggered discretization, there 
% are only 2 tastes.
% This makes them suitable to studying a finite baryochemical potential, with 
% an equal up and down quark chemical potential $\mu_u = \mu_d = \frac{\mu_B}{3}$, without the need to resort to rooting.

The breaking of the hypercubic symmetry causes the appearance of 
relevant and marginal operators which are not present in the 
continuum theory, and thus requires the addition of counterterms. It
was shown in Ref.~\cite{Pernici:1994yj} that there is a single 
relevant operator of dimension 3 appearing, the Lorentz-symmetry 
breaking term $i \bar{\psi} \gamma_\alpha \psi$. It was shown in 
Ref.~\cite{Bedaque:2008xs} that such a term cannot be avoided in 
chirally symmetric, minimally doubled actions. Additionally, the 
dimension 4 counterterm $\bar{\psi} \gamma_\alpha D_\alpha \psi$, 
amounting to a renormalization of the fermionic speed of light 
appears \cite{Pernici:1994yj,Bedaque:2008xs, Capitani:2010nn}. 
Because of quark-gluon interactions, the breaking of hypercubic 
symmetry propagates to the gauge sector, and requires the presence of 
a gluonic counterterm $\sum_{\mu \neq \alpha} F_{\alpha \mu}^2$, of 
dimension 4, which amounts to a renormalization of the gluonic speed 
of light \cite{Pernici:1994yj,Bedaque:2008xs,Capitani:2010nn}.

The two fermionic counterterms (of dimensions 3 and 4) and one 
gluonic counterterm (of dimension 4) can be defined on the lattice as:
\begin{equation}
\begin{aligned}
S^{3f} &= c \sum_x \bar{\psi}(x) i \gamma_\alpha \psi(x)\rm{,} \\
S^{4f} &= (\xi_0-1) \sum_x \bar{\psi}(x) \frac{1}{2} \gamma_\alpha \left( U_\alpha(x)\psi(x+\hat{\alpha}) \right. \\
 & \qquad \qquad \qquad- U_\alpha^\dagger(x-\hat{\alpha})\psi(x-\hat{\alpha}) \!\! \left. \right) \rm{,} \\
S^{4g} &= d_G \sum_x \sum_{\mu\ne\alpha} \operatorname{Re} \operatorname{Tr} \left( 1 - \mathcal{P}_{\mu\alpha}(x)\right)\rm{,}
\end{aligned}
\label{eq:ct}
\end{equation}
where $\mathcal{P}_{\mu \alpha}(x)$ are the plaquettes on the $\mu$-$\alpha$ plane at lattice site $x$. 
Throughout this work, we identify the direction $\alpha$ with the temporal 
direction $\alpha=0$. The non-perturbative tuning of the parameters $c$, $\xi_0$ 
and $d_G$ using mesonic correlation functions was demonstrated in quenched 
QCD in Ref.~\cite{Weber:2015hib}.

The Dirac operator reads:
\begin{align}
D_{\rm KW} \psi(x) \equiv& 
\frac12 \sum_\mu \left[c_\mu(x)~\Gamma_\mu~U_\mu(x)~\psi(x+\hmu) 
\right.\\\nonumber 
& \qquad \left.- c^{-1}_\mu(x-\hmu)~ \Gamma_\mu^\dagger U^\dagger_\mu(x-\hmu)\psi(x-\hmu)\right]\\\nonumber
&+(m + i(3\zeta+c)\gamma_{0}) \psi(x) \, \, ,
\end{align}
where:
\begin{equation}
\Gamma_\mu = \begin{cases}
\gamma_\mu - i \zeta\gamma_{0} & \mu \neq 0\\
\xi_0 \gamma_\mu & \mu=0
\end{cases} \, \, ,
\end{equation}
and the boundary conditions, together with a (in general complex) chemical potential $\mu_q$ are encoded in:
\begin{equation}
c_\mu(x)=e^{\frac{\mu_q}{N_t}~\delta_{0 \mu}}\Phi(x) \, \, ,
\end{equation}
and $\Phi(x)=\pm1$.

The unimproved action has been subject to a detailed discussion in the
PhD thesis of Johannes Weber \cite{Weber:2015oqf}. It was pointed out
that the Karsten-Wilczek term in the action is odd both with respect to charge conjugation
and to the reversal of the $\alpha$ direction, but even for the product of the two.
This term is the only source of $\mathcal{O}(a)$ errors. The sign of the symmetry
breaking contributions can, in principle, be flipped if $\zeta$ is turned from
+1 to -1. This feature can, under some circumstances, be exploited to cancel
the $\mathcal{O}(a)$ effects. In the next section we will follow the
idea of Hamber and Wu \cite{Hamber:1983qa} and address the dimensionality of
the Karsten-Wilczek term.

%%%%%%%%%%%%%%%%%%%%%%%%%%%%%%%%%%%%%%%%%%%%%%%%%%%%%%%%%%%%%%%%%%%%%%%%%%%%%%%%
\begin{figure}
    \centering
    \includegraphics[width=0.45\textwidth]{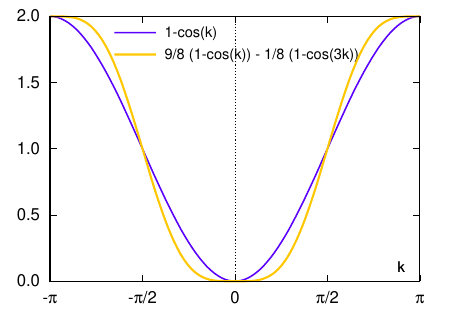}
    \caption{
    The Laplacian term and its 3-hop improved form
    in momentum space in one of the spatial directions.
    The improved operator is proportional to $\sim k^4$ instead of the original $\sim k^2$ form.
    }
    \label{fig:naik_momentum}
\end{figure}
\section{\label{sec:naik}Naik improvement}

In order to reduce discretization errors and facilitate the extrapolation to 
the continuum, in this work we explore the introduction of additional terms 
to improve the action. A simple choice is the Naik 
improvement~\cite{Naik:1986bn}, which we apply in the three spatial 
directions to address the discretization errors coming from the Karsten-Wilczek term.
The Naik term, which is part of the HISQ action, is often used to improve the $\mathcal{O}(a^2)$
errors of the Nabla operator \cite{Follana:2006rc}.

In this work we do not intend to reduce $\mathcal{O}(a^2)$ errors. We also foresee the application
of the Karsten-Wilczek actions in a thermodynamic context and shall maintain
the possibility to apply the reduced matrix formalism. In that formalism multi-hop terms
in Euclidean time are very difficult to accommodate, but a more sophisticated spatial structure 
is not an issue. Thus, we apply the 3-hop term in space only.
Because the action is anisotropic by construction, the lack of
improvement in the temporal direction might be compensated by a smaller
temporal lattice spacing, i.e. a renormalized anisotropy $\xi_f=a_s/a_t>1$, a possibility
that we leave for future work.

In momentum space, like in the Wilson action, the correlator of the KW Dirac operator consists of a Nabla and a Laplacian. The unimproved inverse 
propagator reads:
\begin{equation}
    D_{\rm KW}(k) = \frac{i}{a} \left[
\sum_{\mu=0}^3 \gamma_\mu \xi_\mu \sin\,ak_\mu +\zeta\gamma_0 \sum_{\mu =1 }^3(1-\cos\, ak_\mu) \right] \, \, ,
\end{equation}
where 
\begin{equation}
\xi_\mu = \begin{cases}
1 & \mu \neq 0\\
\xi_0 & \mu=0
\end{cases} \, \, ,
\end{equation}
the $\sin\,a k_\mu$ term represents the single-hop Nabla, and the 
$(1-\cos\,ak_\mu)$ is a Laplacian term in direction $\mu$.

We improve the Dirac operator with extra hop terms with an odd number of steps so that 
we do not spoil the even-odd preconditioning. The cheapest option seems to 
be the Naik improvement with linear triple hops in each spatial direction. The 
improvement of the Nabla is straightforward:
\begin{align} \nonumber
\nabla^{\rm NAIK} \Psi &=  \frac{i}{a}\left[s_1 \sin\,a k_j + s_3\sin\,3a k_j\right]\Psi(k) \\
&= i\left[ k_j + 0 k_j^3 +\mathcal{O}(k^5)\right] \, \, ,
\end{align}
where the requirement in the last equality is satisfied with the choice
$s_1=9/8$, $s_3=-1/24$.

With the improvement of the Laplacian the goal is to suppress the momentum 
dependence of the cosine term altogether, simultaneously lifting the momenta 
in the doubler's part of the Brillouin zone. 
In momentum space we can eliminate the $\mathcal{O}(k^2)$ term in the 
expression
\begin{equation}
(1-\cos\,a k_j) \to c_1 (1-\cos\,a k_j) + c_3 (1-\cos\,3a k_j)
\end{equation}
with the choice $c_1=9/8$, $c_3=-1/8$.

The resulting Laplacian operator in one spatial direction, in 
momentum space, is shown in Fig.~\ref{fig:naik_momentum}. We can appreciate how the improvement of the Laplacian reduces the 
momentum dependence around $k=0$, and at the same time lifts the momenta in 
the vicinity of the edges of the Brillouin zone, as desired.

\begin{widetext}

The full 3D Naik-improved Dirac operator reads:
\begin{align}\label{improvedD}
D^{\rm NAIK} \psi(x) \equiv& 
\frac12 \sum_\mu \left[
c_\mu(x)~\Gamma^{\rm (1)}_\mu~U_\mu(x)~\psi(x+\hmu) 
- c^{-1}_\mu(x-\hmu)~ \Gamma^{\rm(1)\dagger}_{\mu} U^\dagger_\mu(x-\hmu)\psi(x-\hmu) \right. \\\notag
& \qquad
+c_\mu(x)c_\mu(x+\hmu)c_\mu(x+2\hmu)~\Gamma^{\rm (3)}_\mu~
U_\mu(x)U_\mu(x+\hmu)U_\mu(x+2\hmu)
~\psi(x+3\hmu) \\\notag
& \qquad
\left.
- c^{-1}_\mu(x-\hmu)c^{-1}_\mu(x-2\hmu)c^{-1}_\mu(x-3\hmu)~ \Gamma^{\rm(3)\dagger}_{\mu} 
U^\dagger_\mu(x-\hmu)U^\dagger_\mu(x-2\hmu)U^\dagger_\mu(x-3\hmu)
\psi(x-3\hmu)\right]
\\\notag
&+(m + i(3\zeta+c)\gamma_0)~\psi(x) \, \, ,
\end{align}
\end{widetext}
where we used the new notations:
\begin{align}
\Gamma_\mu^{(1)} &= \begin{cases}
\bar s_1 \gamma_\mu - i \bar c_1 \gamma_{0} & \mu \neq 0\\
\xi_0\gamma_0 & \mu=0
\end{cases} \, \, , \\ \nonumber \\[1pt]
\Gamma_\mu^{(3)} &= \begin{cases}
\bar s_3 \gamma_\mu - i \bar c_3 \gamma_{0} & \mu \neq 0\\
0 & \mu=0
\end{cases} \, \, ,
\end{align}
and the additional parameters satisfy
\begin{align}
1 & = \bar s_1 + 3\bar s_3 \, \, ,\\
\zeta &=\bar c_1 + \bar c_3 \, \, .
\end{align}
In the improved Laplacian we have:
\begin{align}
\bar c_1 = \zeta \frac{9}{8} \quad\mathrm{and}\quad
\bar c_3=-\zeta \frac{1}{8}\,,
\label{eq:c_naik_req}
\end{align}
and in the improved Nabla operator we need:
\begin{align}
\bar s_1 = \frac{9}{8} \quad\mathrm{and}\quad
\bar s_3=-\frac{1}{24} \, \, .
\end{align}

In general, with the new gamma matrix structure the half-vector-trick
cannot be used. However, it becomes applicable if 
$\Gamma^j\sim\gamma^j\mp i\gamma^0$. Hence, one may sacrifice the exact Naik 
(spatial) tree-level improvement to utilize the half-vector trick. We note 
that this is not a complete tree-level improvement, as the temporal direction
remains unimproved. In fact, the main goal of this improvement program
is to minimize the lattice artefacts brought in by the 
$a\zeta\gamma^0\triangle$ operator, through replacing the Laplacian by
a higher order derivative, while the improvement of the spatial Nabla is 
of secondary importance.
Thus, we have the freedom to fix $\bar s_1$, $\bar s_3$ and $\zeta$ 
to maintain $\bar s_1 = \bar c_1$ and $\bar s_3=\bar c_3$, 
while also keeping $\bar c_1 = \zeta 9/8$ and $\bar c_3=-\zeta/8$. 
This can be achived with the choice:
\begin{align}
    \bar s_1&=\bar c_1=1.5 \, \, ,\nonumber\\
    \bar s_3&=\bar c_3=-1/6 \, \, ,\nonumber\\
    \zeta & =4/3 \, \, .
    \label{eq:naik_coeff_hvt}
\end{align}

This is the choice we employ in this work. The case of no improvement 
means $\bar c_3 = \bar s_3 = 0 $, which requires $c_1=1$, so 
$\bar c_1=\zeta$ and $s_1=\bar s_1=1$.
This admits the half vector trick if and only if $\zeta=\pm 1$.

%%%%%%%%%%%%%%%%%%%%%%%%%%%%%%%%%%%%%%%%%%%%%%%%%%%%%%%%%%%%%%%%%%%%%%%%%%%%%%%%
\section{\label{sec:renorm}Non-perturbative renormalization}

\subsection{Lattice details}

% \textbf{Subsection a: Include lattice details: size, preconditioning, half-vector trick, solvers, etc.}
% \newline
This is a mixed action study. We use five gauge ensembles that were
generated using the 4stout staggered action with tree-level Symanzik
improvement in the gauge sector \cite{Bellwied:2015lba}.
The ensembles were generated along the line of constant physics
with 2+1+1 flavors, where the pion and kaon masses were tuned
to assume the physical ratios to $f_\pi$. 
The ensembles in this work are listed in Table \ref{tab:ensembles}.

\begin{table}
\begin{center}
\begin{tabular}{|c|c|c|c|c|}
\hline
$\quad \beta \quad$ & $\quad a$ [fm] $\quad$ & 
$\quad N_x\times N_\tau \quad $ &
$\quad m_{ud} \quad $ &
$\quad m_s \quad $ 
\\
\hline
3.6376& 0.1581 &$40\times64$ &	0.00243956	&0.0687299\\
3.7089& 0.1308 &$48\times64$ &	0.00201959	&0.0566136\\
3.7589& 0.1152 &$64\times96$ &	0.00178099	&0.0496697\\
3.8360& 0.0953 &$64\times96$ &	0.00147176	&0.0407946\\
4.0126& 0.0638 &$96\times144$&0.000958973&	0.0264999\\
\hline 
\end{tabular} 
    \caption{\label{tab:ensembles}
    The staggered ensembles and the bare parameters of
    the 4stout action. The charm mass is always $m_c=11.85 \, m_s$. Each ensemble is 144 configurations strong.
    }
\end{center}
\end{table}

\begin{figure} \centering 
    \includegraphics[width=.5\textwidth]{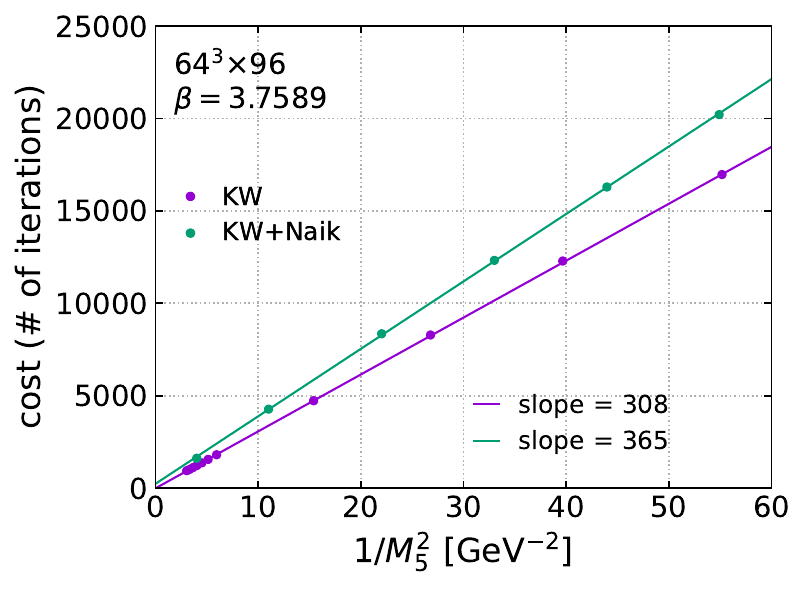}
    \caption{Number of iterations vs. the pion mass. The improved
    action has a slightly higher condition number.
    }
    \label{fig:cost} 
\end{figure}

Throughout this work we employ the $w_0$ based scale setting with $w_0=0.17236$~fm, as of Ref.~\cite{Borsanyi:2022ygn}.
All masses, expressed in MeV, are based on this $w_0$ scale obtained on the 4stout-staggered gauge configurations. 

We implemented the Karsten-Wilczek Dirac operator both for the
standard and the 3-hop setup. Having avoided a two-hop form
for the improvement we could benefit from even-odd preconditioning
\cite{Luscher:2010ae}. We note that, being chiral, the Karsten-Wilczek 
operator turns an $L$-handed Weyl spinor into an $R$-handed spinor
and vice versa. This allows one to formulate a solver for the $L$-spinor,
with the benefit of working with a half-sized spinor. This trick is the
analogous of the even-odd decomposition of the staggered operator.

The solver, which is based on the $L/R$-decomposition, is
outperformed by the even-odd preconditioned solver, because the latter works
with an effective operator with reduced condition number. Indeed,
the Karsten-Wilczek operator lends itself to such preconditioning:

\begin{equation}
\begin{pmatrix}
M_{ee} & M_{eo}\\
M_{oe} & M_{oo}
\end{pmatrix}
\begin{pmatrix}
\phi_e \\
\phi_o 
\end{pmatrix}
=
\begin{pmatrix}
\eta_e \\
\eta_o 
\end{pmatrix}
\end{equation}
for some source $\eta$ and solution $\phi$.
A key feature of the Karsten-Wilczek operator is that 
$M_{ee}$ and $M_{oo}$ are strictly local.
One can express $\phi_o$ in terms of $\phi_e$:
\begin{equation*}
    \phi_o = M_{oo}^{-1} \left(\eta_o - M_{oe} \phi_e \right).
\end{equation*}
Therefore one can solve for $\phi_e$ alone:
\begin{equation*}
    (M_{ee} - M_{eo}M_{oo}^{-1} M_{oe} )\phi_e\equiv \hat{M} \phi_e = \eta_e - M_{eo} M_{oo}^{-1}\eta_o 
\end{equation*}
in which $\hat{M}$ is even-even only. $\gamma_5\hat{M}$ is hermitian, though not
positive. This is in contrast with the $L/R$ decomposed solver where the product
$M_{LR}M_{RL}$ is hermitian and positive, thus a conjugate gradient algorithm is
applicable. Luckily the conjugate residual algorithm (see e.g. Ref
~\cite{Saad_2003}) performs well for $\gamma_5\hat{M}$. 
The so called ``half vector trick'', 
that relies on the projector nature of the gamma matrix structure of the non-local terms, is an independent optimization we adopted. In total, the 3-hop operators
arithmetic weight is roughly double that of the Wilson action. We plot the iteration
count as a function of the pseudoscalar mass for one of our ensembles
in Fig.~\ref{fig:cost}.

\subsection{Tuning}

In this section we illustrate the procedure we employed to tune 
the parameters of the KW action. We need to carry out a 
multi-dimensional tuning involving the KW parameters $c$ and $\xi_0$, 
as well as the fermion mass $m_0$. The KW parameter $d_G$ is not
tuned, as the gauge configurations were generated with our 4stout 
staggered action.
We will show in the following that the tuning largely decouples, and
the three parameters $c$, $\xi_0$ and $m_0$ can be tuned one at a time.

\subsubsection{Tuning the dimension-three counterterm}

As shown in Ref.~\cite{Weber:2015hib}, in order to tune $c$ it is 
convenient to exploit the existence of oscillating contributions, 
related to fermion doubling \cite{Weber:2015oqf}, to the correlation 
functions in the direction $\alpha$ of certain meson interpolating 
operators 
\begin{equation}C_\Gamma(n-m) \sim \left\langle \bar{\psi}(n) \Gamma \psi(n) \bar{\psi}(m) \Gamma \psi(m) \right\rangle \, \, .
\end{equation} 
% where $\Gamma$ is some combination of $\gamma$-matrices. 
Crucially, the frequency of these oscillations is very sensitive to 
the value of $c$. 

\begin{figure}
 \centering 
\includegraphics[trim=5 0 40 15, clip, width=.45\textwidth]{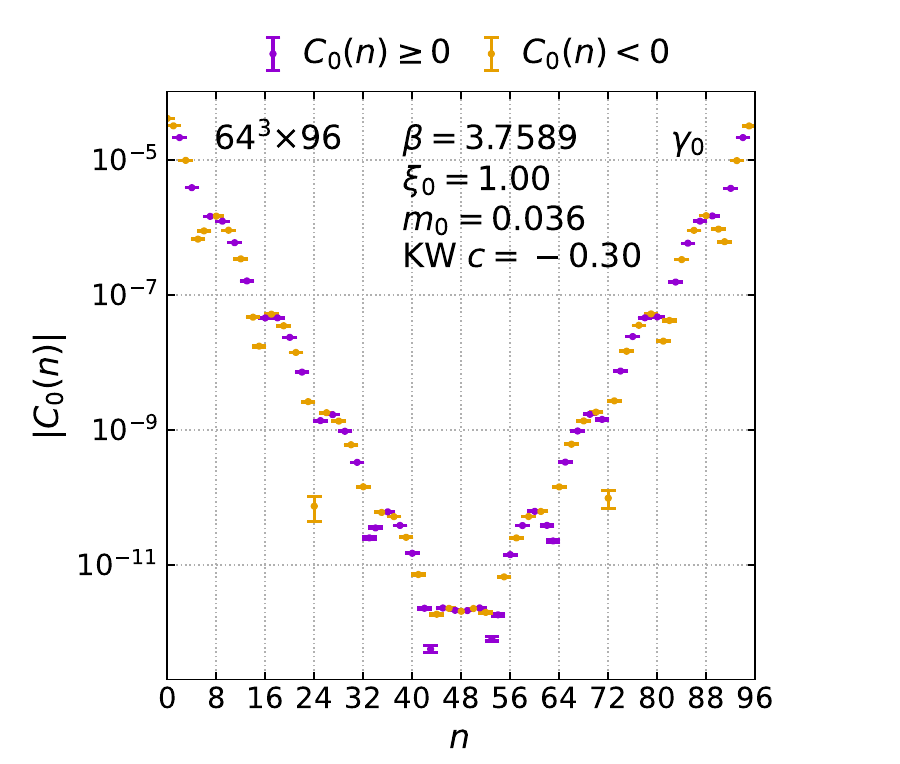} 
\includegraphics[trim=5 0 40 15, clip, width=.45\textwidth]{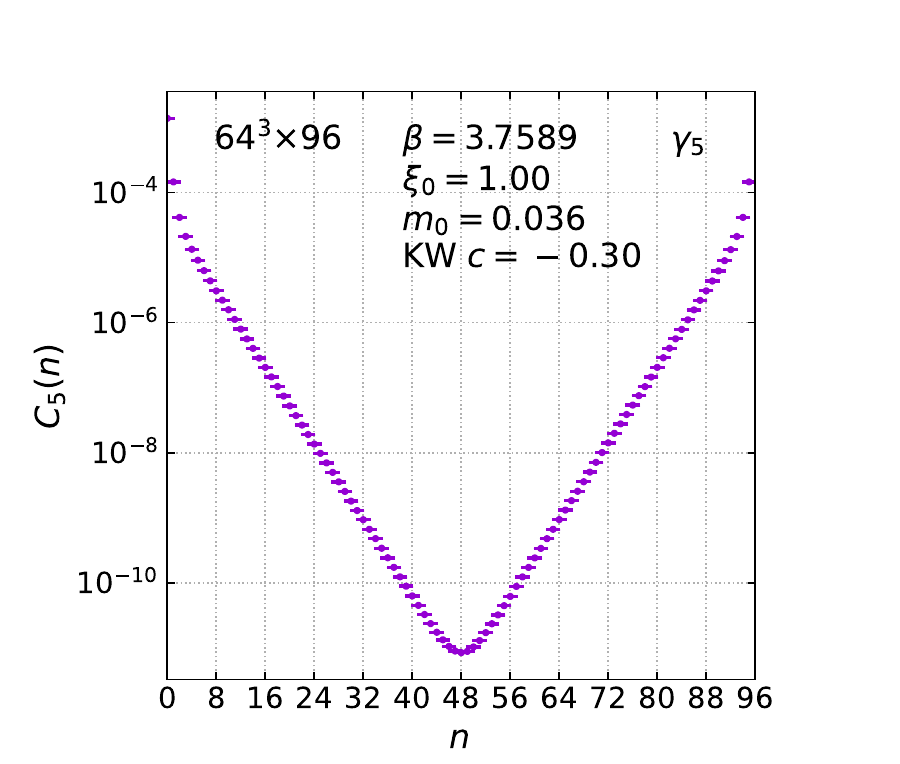}
\caption{Correlators for the $\gamma_0$ (top) and $\gamma_5$ (bottom) fermion
channels for the Karsten-Wilczek action. The frequency of the oscillating
$\gamma_0$ correlator depends on the $c$ parameter of the action. A beat
oscillation on top of the $(-1)^n$ oscillation (visible in the top plot)
occurs when the $c$ parameter is detuned. \label{fig:correlators} }
\end{figure}

Fermion parity partners can be identified by the spin-taste structure 
of the KW action, which was thoroughly discussed in 
Refs.~\cite{Weber:2016dgo,Weber:2023kth}.  
For pseudoscalar mesons, two relevant tastes are the $\gamma_0$ 
(since $\alpha=0$ in this work) and $\gamma_5$ channels, where the 
latter can be identified with the physical pion. The correlators for 
$\gamma_0$ and $\gamma_5$ are taken \textit{parallel} to the 
preferred axis of the KW action, $\alpha=0$ in this case. 
The parallel correlator for $\gamma_0$ exhibits oscillations while that of 
$\gamma_5$ does not. The tuning criterion for the $c$ parameter is where the 
frequency spectrum of the oscillations of the $\gamma_0$ channel recovers 
its tree-level form \cite{Weber:2015oqf}. At the tuned $c$, the oscillation 
of the $\gamma_0$ correlator is described by $(-1)^n$, where $n$ is the temporal separation
of source and sink.
The frequency of the 
oscillation at a given $c$ value can be described by 
$\omega = \omega_c + \pi$, where $\pi$ is the frequency of the rapid 
oscillations at tuned $c$ given by the $(-1)^n$ factor, and $\omega_c$ is a 
beat frequency that appears in the rapid oscillations when $c$ is detuned.
Thus, the tuning criterion can be equivalently stated as where 
$\omega_c = 0$ and the beat vanishes~\cite{Weber:2015oqf}. 
Fig.~\ref{fig:correlators} shows an example of correlators in the $\gamma_0$ 
and $\gamma_5$ channels at detuned $c$, where both the rapid oscillation and 
the beat oscillation can be observed for $\gamma_0$. 

We average over the symmetric halves of the $\gamma_0$ correlator 
$C_0(n)$ and eliminate the rapid oscillation with a factor of 
$(-1)^n$,
\begin{equation}
C(n) = (-1)^n\frac{1}{2}(C_0(n) + C_0(N_t-n)),
\end{equation} 
for $0 \leq n \leq N_t/2$, where $N_t$ is the temporal extent of the 
lattice. Then, the correlator is well described by the model 
\begin{equation}
C(n) \approx A \cosh(M(n-N_t/2)) \cos(\omega_c n - \phi) 
\end{equation}
where $M$ is the mass of $\gamma_0$ in lattice units and $\phi$ is a 
phase factor, and one can see that the beat frequency and the mass 
decouple. 
% We use various fitting methods for extracting $M$ and $\omega_c$ 
% because of the exotic shape of the correlator. 

\begin{figure}
 \centering 
\includegraphics[width=.42\textwidth]{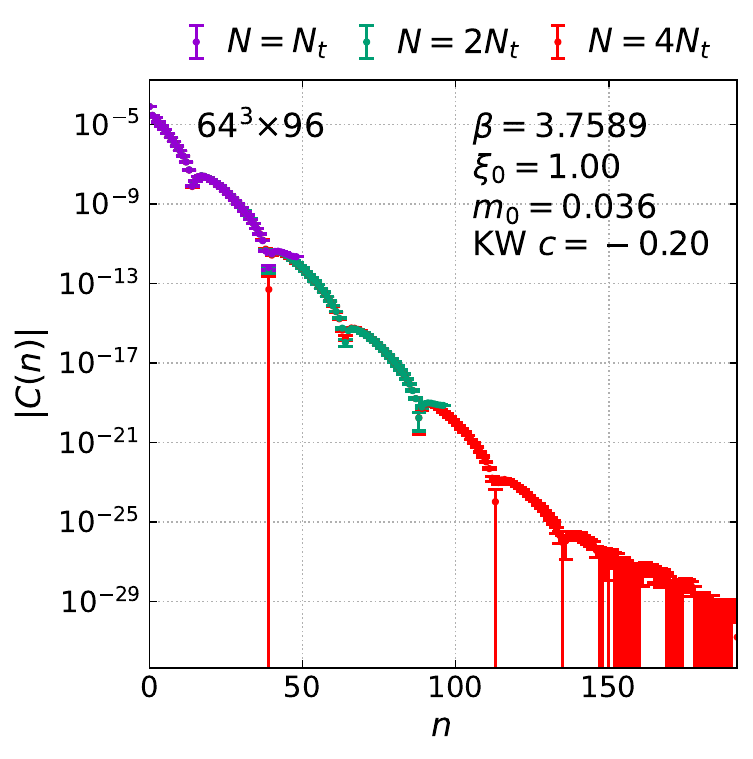}
\includegraphics[width=.42\textwidth]{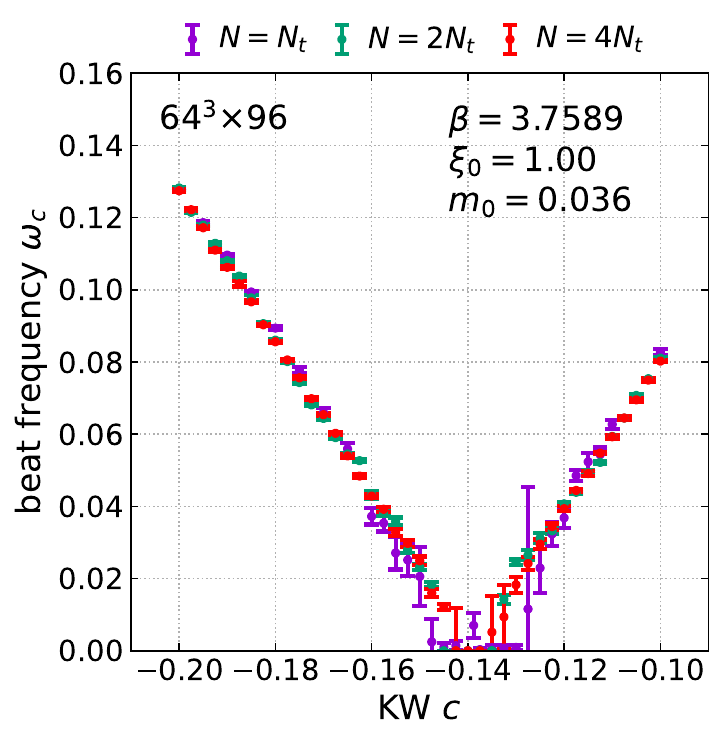}
\caption{(Top) The symmetrized $\gamma_0$ correlator measured with different
amounts of tiling with stored gauge configurations. The proper use of tiling
permits longer correlation lengths to be measured and increases the precision
of the measurement of the beat frequency $\omega_c$. (Bottom) Measurements of
$\omega_c$ as a function of the $c$ parameter at the same three tilings.
\label{fig:tiling} }
\end{figure}

An immediate difficulty that arises in tuning $c$ is due to the finite 
length of the lattice, which makes fitting the frequency unreliable 
when the wavelength of the beat is greater than $N_t$, in the 
relative vicinity of tuned $c$. 
This can be ameliorated through the use of tiling gauge 
configurations. Tiling means the gauge configurations are 
concatenated in the desired direction (time in our case) to form 
$N_x^3{\times}(2 N_t)$ or $N_x^3{\times}(4 N_t)$ lattices on which 
the propagator is measured. The top panel of Fig.~\ref{fig:tiling} 
shows an example of $C(n)$ at detuned $c$ for 1${\times}$ (i.e., no 
tiling), 2${\times}$ and 4${\times}$ tiling.
This allows to study the propagator in a longer range 
than that of the dynamical simulation. Tiling extends the idea of a 
mixed action study where the quarks live on an extended lattice 
without maintaining the long wave-length gauge fluctuations, which 
would anyway be irrelevant for the divergent parts of the Feynman 
diagrams. Thus, when used properly, tiling the stored gauge 
configurations increases the precision with which $\omega_c$ is 
measured. This is evident from the bottom panel of 
Fig.~\ref{fig:tiling}, where the measured beat frequency 
$\omega_c(c)$ is shown for each level of tiling. We use 
4${\times}$ and occasionally 1${\times}$ tiling throughout this 
study.

\begin{figure}
 \centering 
\includegraphics[width=.505\textwidth]{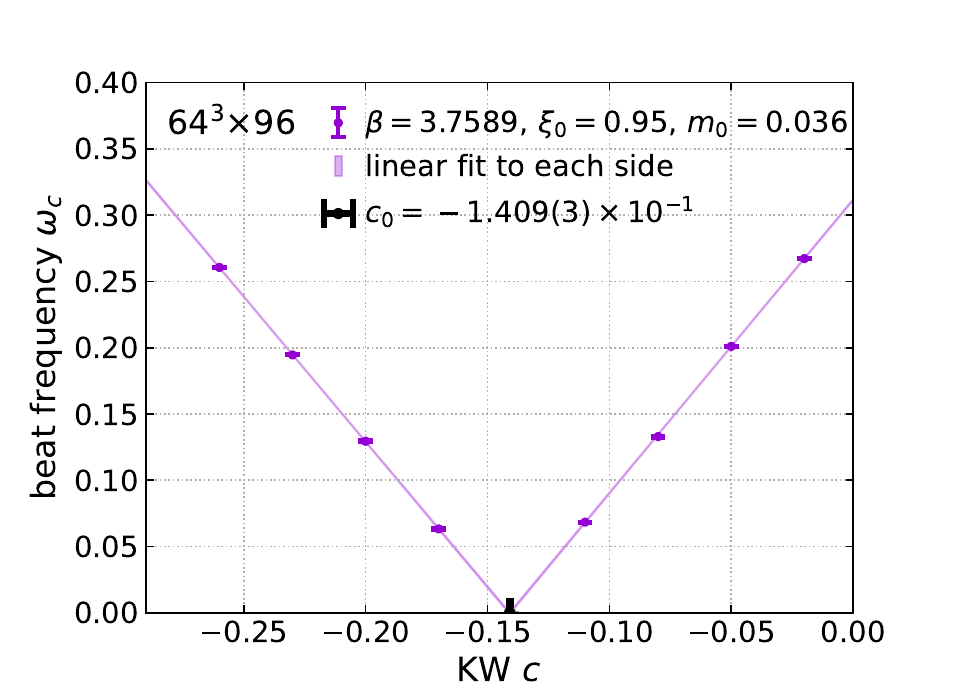} \caption{The beat frequency $\omega_c$ as a function of the $c$ 
parameter. The $c$ parameter is tuned at the value where the beat 
frequency vanishes. \label{fig:c_tuning} }
\end{figure} 

Regardless of tiling, $c$ is tuned for a given pair $\xi_0, m_0$ 
by scanning through $c$ and searching for $\omega_c(c)=0$. As
shown already in Ref.~\cite{Weber:2015hib}, the behavior of 
$\omega_c(c)$ is basically linear on both sides of the tuned value 
of $c$. An example of this is shown in Fig.~\ref{fig:c_tuning}, 
where the final result is obtained by combining linear fits on 
either side of the tuned $c$. Throughout this work, we find that linear fits to both sides were unproblematic, and yielded consistently good p-values. 

\subsubsection{Tuning the bare anisotropy $\xi_0$}

The physical anisotropy can be determined using a 
$\gamma_5$-correlator in a direction perpendicular to $\alpha$.
It is defined as the ratio of the perpendicular (spatial) mass of 
$\gamma_5$ to the parallel (temporal) mass:
$\xi_f = M_\perp/M_\parallel$. This allows the tuning of the bare 
anisotropy $\xi_0$ appearing in the action.

\begin{figure} \centering
\includegraphics[width=.46\textwidth]{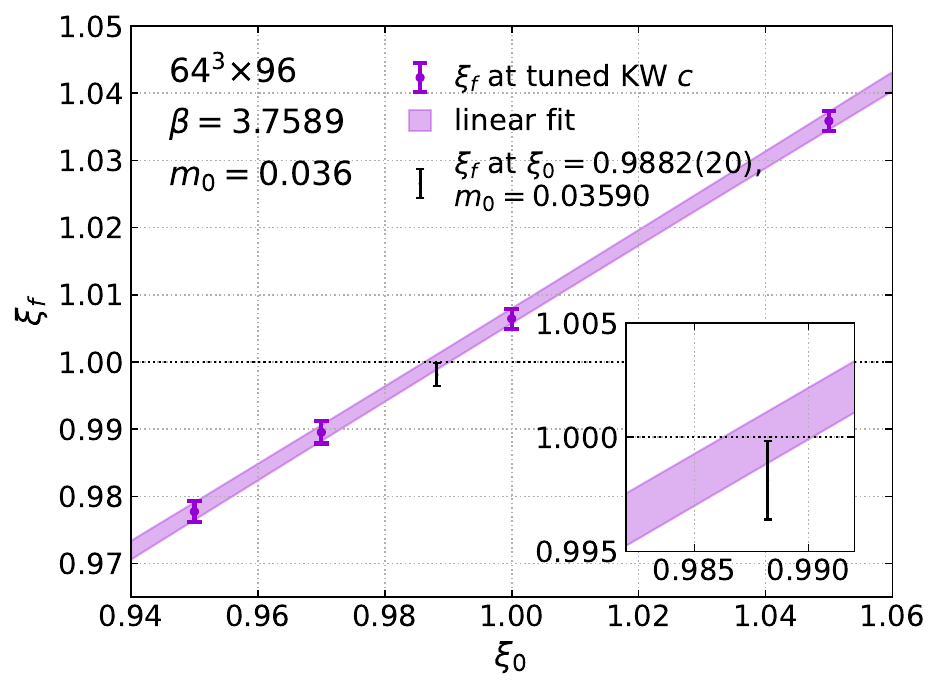}
\includegraphics[width=.48\textwidth]{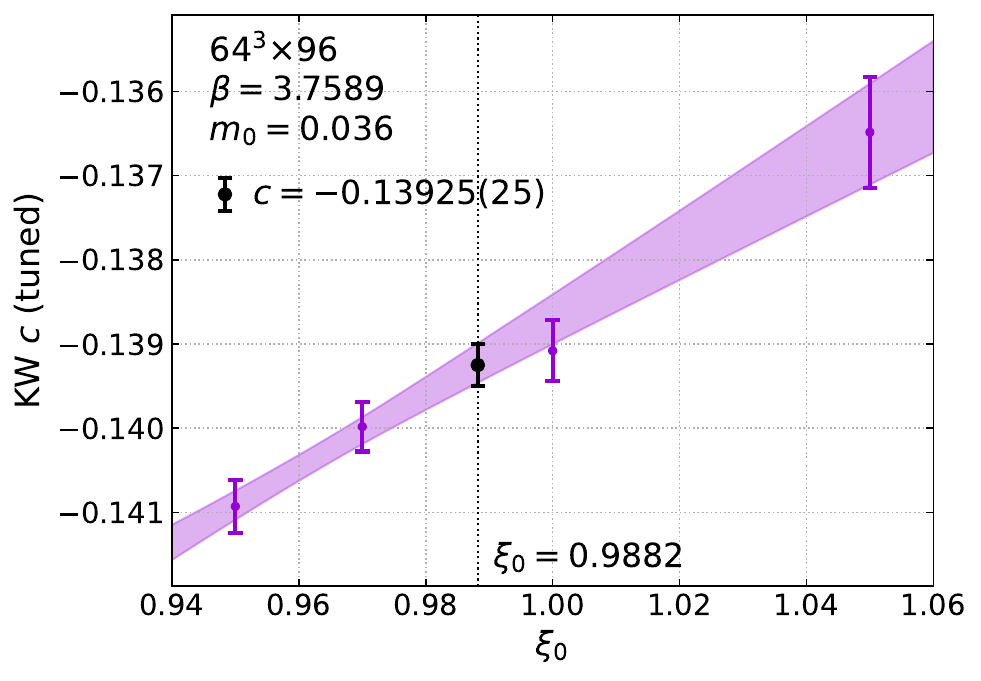} 
\caption{(Top) The renormalized fermion anisotropy 
$\xi_f = M_\perp/M_\parallel$ as a function of the bare anisotropy 
$\xi_0$. The $c$ parameter is individually tuned at each $\xi_0$ 
value. The tuning criterion for $\xi_0$ is $\xi_f = 1$. 
(Bottom) Interpolation to the tuned value of the $c$ parameter at 
the tuned value of $\xi_0$. \label{fig:c_xi_tuning} }
\end{figure}

Following the procedure shown in Fig.~\ref{fig:c_tuning},
we tune $c$ for every value of $\xi_0$. Then, $\xi_0$ is tuned by 
imposing the desired renormalized anisotropy $\xi_f$. Since we 
perform measurements on isotropic staggered configurations, we tune 
by interpolating $\xi_f(\xi_0)$ at tuned $c$ to $\xi_f(\xi_0)=1$. 
An example of this is shown in the top panel of 
Fig.~\ref{fig:c_xi_tuning}. 

At this point, it is necessary to simultaneously tune $c$ and 
$\xi_0$. Fortunately, the tuned value of $c$ depends only mildly on 
the tuned anisotropy $\xi_0$. Hence, with an interpolation of 
$c(\xi_0)$ to the tuned value $\xi_0$, we determine the final tuned 
$c,\xi_0$ pair (at fixed $m_0$). The bottom panel of
Fig.~\ref{fig:c_xi_tuning} shows an example of this interpolation.
In the top panel of Fig.~\ref{fig:c_xi_tuning} we also show with a
black point the renormalized anisotropy $\xi_f$ from a dedicated run
at the final tuned values of $c$ and $\xi_0$, which shows a result 
compatible with 1, hence confirming the validity of our 
procedure.

\begin{figure}
 \centering 
\includegraphics[trim=0 0 0 0, clip, width=.5\textwidth]{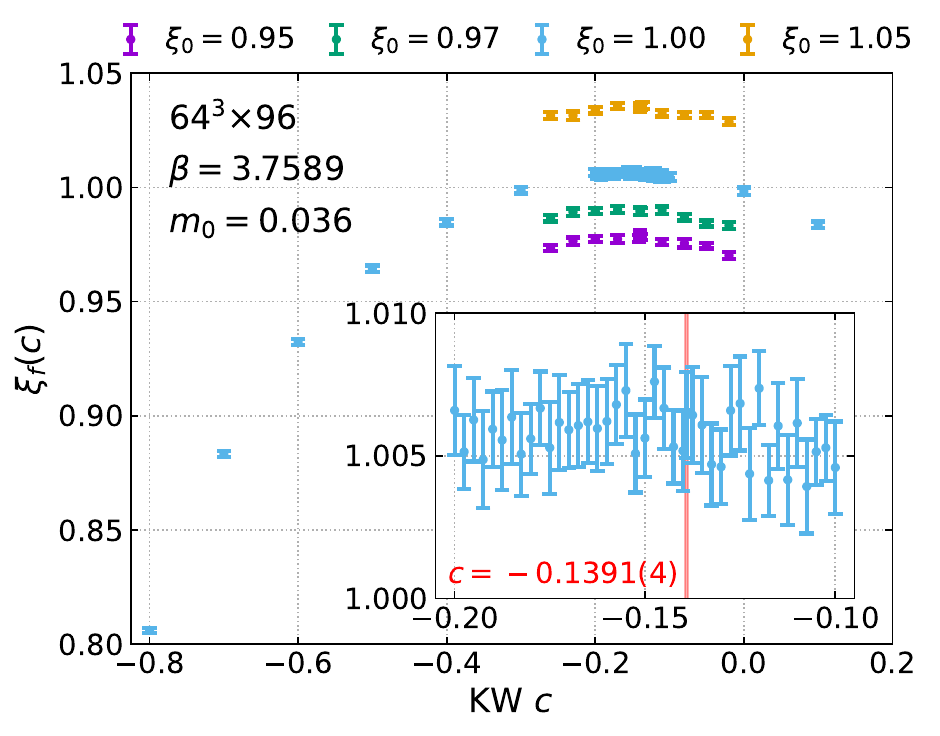}
 \caption{The renormalized anisotropy $\xi_f$ as a function of the $c$ parameter at various fixed bare anisotropy values $\xi_0$. 
 $\xi_f$ reaches a maximum around tuned $c$, and therefore it is stable against slight mistunings of $c$.\label{fig:renorm_xi_vs_c} } 
\end{figure}

Finally, it is reasonable to ask how accurately one needs to tune $c$ 
and $\xi_0$, or how stable the result is to slight mistunings. 
Thankfully, the results are quite stable, as the $\xi_f(c)$ function 
has vanishing derivative (a maximum) near the tuned value. This is 
shown in Fig.~\ref{fig:renorm_xi_vs_c}. 

This is not surprising: the anisotropy parameter does not break $C$ or $T$, i.e.
it's even in $\zeta$, while a de-tuning of $c$ introduces an odd effect in 
$\zeta$. 
Thus, $d\xi/dc$ is odd and is expected to
vanish up to higher orders in the lattice spacing, if the counterterm is
correctly tuned. A similar behavior is observed with the mass of the pseudoscalar.

\input{TableI.tex}    

\subsubsection{Tuning the bare mass $m_0$}

\begin{figure}
\centering 
\includegraphics[trim=20 0 80 0, clip, width=.45\textwidth]{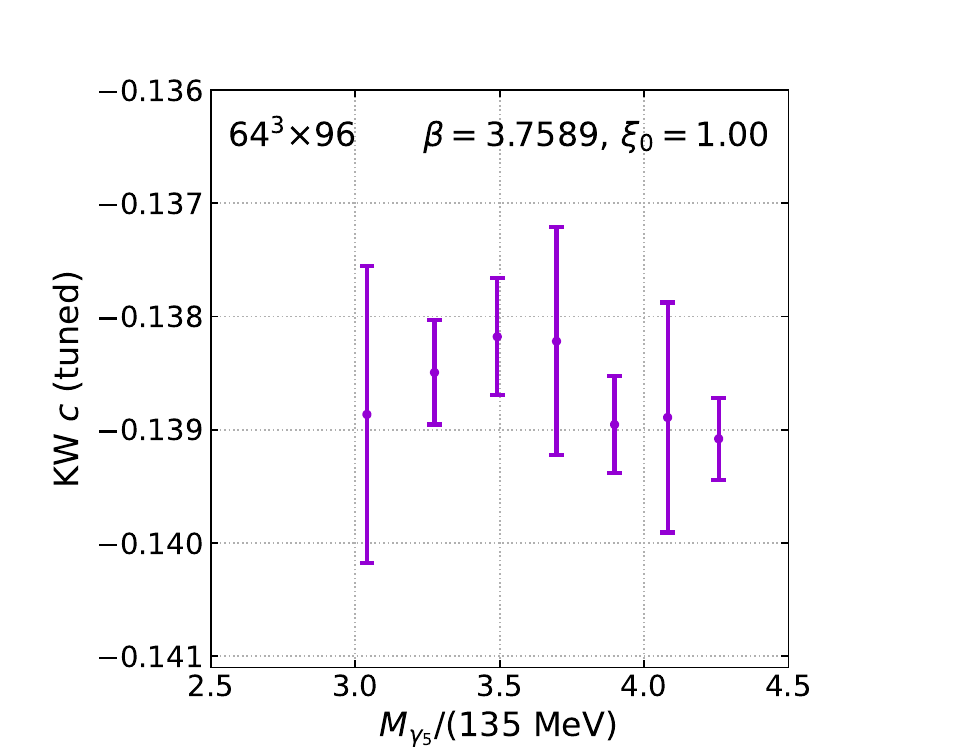} 
\includegraphics[trim=20 0 80 0, clip, width=.45\textwidth]{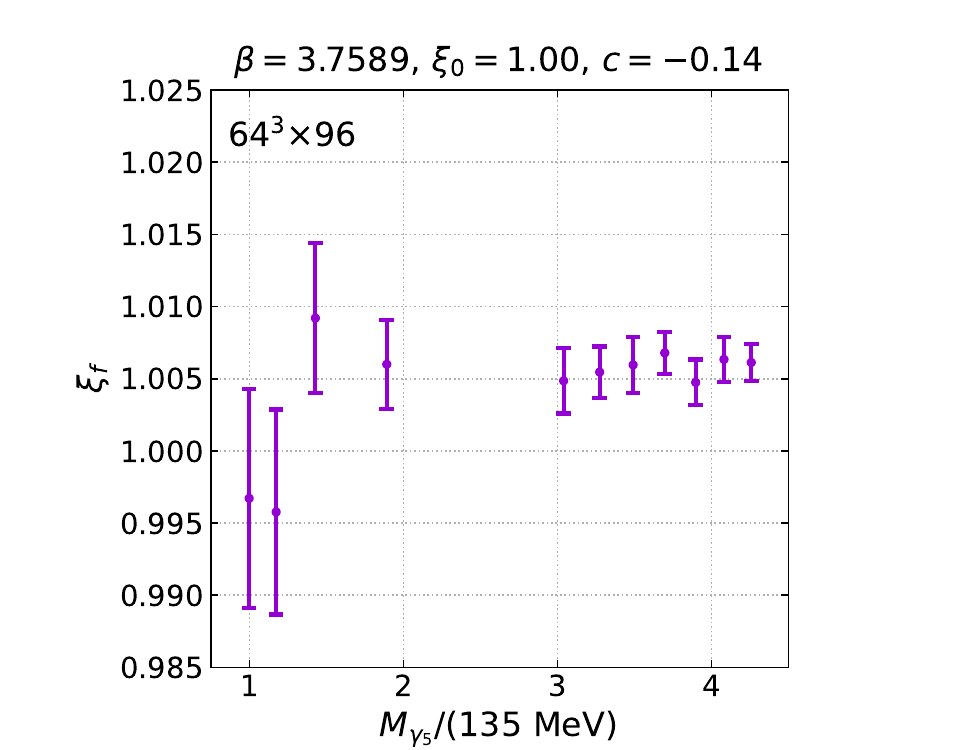} 
\caption{Stability of the tuned $c$ value at fixed $\xi_0$ (top) and 
the renormalized anisotropy $\xi_f$ at fixed $\xi_0$ and $c$ 
(bottom) with respect to the pseudoscalar mass $M_{\gamma_5}$. 
$\xi_f$ remains stable even all the way down to the physical pion 
mass $M_\pi \approx 135 \ {\rm MeV}$. The larger error bars at the 
small $M_{\gamma_5}$ values are due to smaller statistics.  
\label{fig:m5_variation} }
\end{figure}

In this work, we carry out the tuning procedure first at 
a value of the pseudoscalar mass $M_{\gamma_5} = 578.4$ MeV. The 
final results of the tuning of $c$ and $\xi_0$ at several lattice 
spacings, with the $\gamma_5$ mass held constant at this value, are 
listed in Tables~\ref{tab:tuned_values_noim} 
and~\ref{tab:tuned_values_impr}. Then, we study the dependence of 
the tuned $c$ and $\xi_0$ values on $M_{\gamma_5}$, down to the 
physical pion mass. We find that the tuned value of $c$ for a given 
$\xi_0$ is very stable with respect to changes in $M_{\gamma_5}$. 
Moreover, for fixed $c$ and $\xi_0$, the renormalized fermionic 
anisotropy $\xi_f$ is also very stable with respect to changes in 
$M_{\gamma_5}$, as its value remains in statistical agreement with 1 
all the way down to the physical pion mass 
$M_{\gamma_5} = M_\pi \approx 135 \ {\rm MeV}$. These results are 
shown in Fig.~\ref{fig:m5_variation}. Finally, 
$M_{\gamma_0}$ and $M_{\gamma_5}$ are concave functions of 
$c$, with a minimum near the tuned value of $c$. They are 
thus generally quite robust against mistunings of $c$, 
especially $M_{\gamma_5}$. Additional details on the 
dependence of $M_{\gamma_0}$ and $M_{\gamma_5}$ on $c$ can 
be found in Ref.~\cite{Godzieba:2024uki}.

\subsubsection{Smearing}

\begin{figure}
 \centering 
\includegraphics[trim=0 0 0 0, clip, width=.45\textwidth]{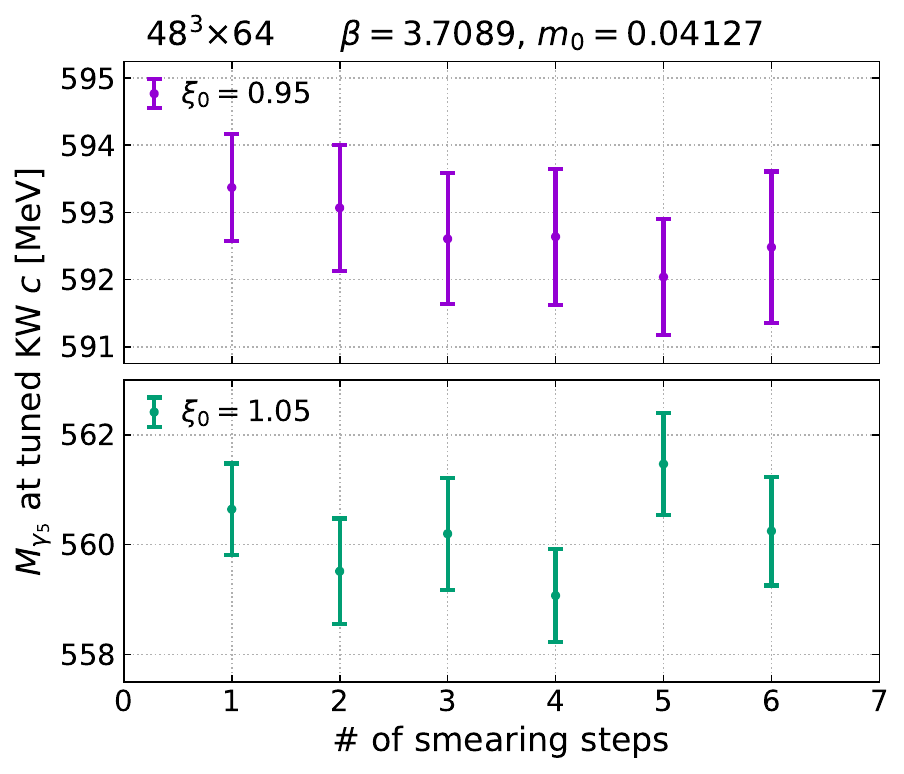} 
\includegraphics[width=.45\textwidth]{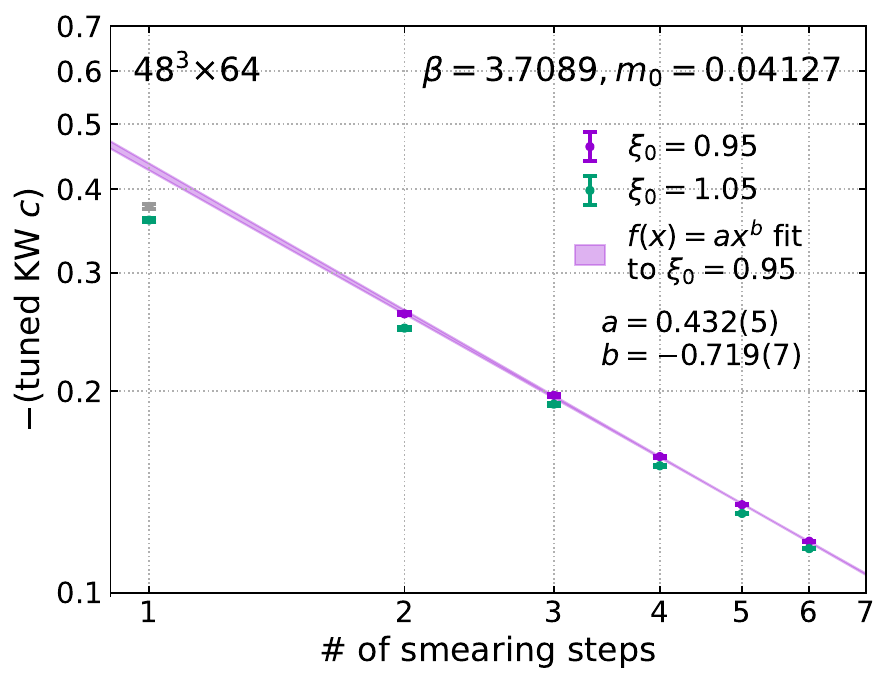}
\caption{ (Top) The pseudoscalar mass 
$M_{\gamma_5}$ as a function of the number of applied smearing 
steps. The number of smearing steps does not visibly affect the 
observed value of the mass.
(Bottom) The variation of the tuned value of $c$ 
for the case of the unimproved action
with the number of stout 
smearing steps applied. We used 4 stout smearing steps throughout 
the rest of the analysis. 
\label{fig:smearing} }
\end{figure} 

Lastly, we consider the effect of different levels of stout 
smearing on the tuning. In princple, each bare parameters
have to be retuned at a new choice of the smearing level.
We observe in the top panel of Fig.~\ref{fig:smearing} that
the actual pseudo-scalar mass ($M_{\gamma_5}$) is within
error independent on the smearing level while keeping there bare mass $m_0$ fixed. This allows us the simplification to
address the variation of the counterterm $c$ with smearing
using the same fixed $m_0$ value.
For also a fixed value of $\xi_0$, we find the effect of 
applying more smearing steps to be a power law decrease in the 
magnitude of the tuned $c$ value. An example of this is shown in the  bottom panel of Fig.~\ref{fig:smearing}. 

Throughout this study,  we will use four steps of stout smearing, the same
smearing level that was used in the staggered simulations to generate the 
ensembles.

\subsubsection{Hierarchy}

To summarize, our tuning procedure proceeds in the following steps:
\begin{enumerate}
    \item tune $c$ at fixed bare anisotropy $\xi_0$, for a given 
    pseudoscalar mass $M_{\gamma_5}$, which can be chosen to be 
    large for numerical convenience;
    \item repeat the procedure for different $\xi_0$ values, tuning 
    $c$ each time, then interpolate to $\xi_f(\xi_0)=1$;
    \item interpolate the tuned $c$ at the tuned $\xi_0$;
    \item finally, tune the bare mass $m_0$ to fix the physical 
    pseudoscalar mass ($c$ and $\xi_0$ don't need to be tuned again).
\end{enumerate}

A hierarchy of the bare parameters is thus established. Most 
critical in the tuning procedure is the $c$ parameter, as the 
physical masses of oscillating fermion channels are highly dependent 
on it. The bare anisotropy $\xi_0$ follows in importance. Finally, 
the bare mass $m_0$ (alternatively the physical mass $M_{\gamma_5}$) 
is last, as it has the mildest effect.

%%%%%%%%%%%%%%%%%%%%%%%%%%%%%%%%%%%%%%%%%%%%%%%%%%%%%%%%%%%%%%%%%%%%%%%%%%%%%%%%%%%
\section{\label{sec:fpi}Pion decay constant} 

In order to test both the unimproved and the Naik improved versions 
of our KW action, we turn to the computation of the pion decay 
constant $f_\pi$, which we continuum extrapolate in both cases. We 
also compare these results with staggered results on the same gauge 
ensembles, for which the quark mass was tuned to obtain the same 
value of $M_{\gamma_5} = 578.4$ MeV. Both KW actions are run with the 
parameters listed in Tables~\ref{tab:tuned_values_noim} 
and~\ref{tab:tuned_values_impr}. The 3-hop action was not evaluated
on the finest lattice due to memory limitations in the current implementation.

The pseudoscalar decay constant $f_\pi$ in the continuum (with the
convention that the physical value is $131.5$ MeV) is defined as 
\begin{equation}
    \langle \pi(p) | A_\mu(x)|0\rangle = e^{ipx} p_\mu f_\pi \, \, ,
\end{equation}
with the continuum PCAC
\begin{equation}
    A_\mu(x) = \bar{\psi}_u(x)\gamma_\mu\gamma_5\psi_d(x) \, \, .
\end{equation}
The corresponding Ward Identity is
\begin{align}
    \partial_\mu A_\mu(x) &= (m_u+m_d) P(x),\\
    P(x) &\equiv \bar{\psi_u}(x) \gamma_5 \psi_d(x), 
\end{align}
where $P(x)$ is the pseudoscalar density, and $m_u$ and $m_d$ are the masses of quark $u$ and $d$, respectively. 
Therefore,
\begin{equation}
    \frac{1}{\sqrt{V_{3d}}}\int_{\vec{x}}\langle \pi(m_\pi,\vec{0}) | P(0,\vec{x})|0\rangle = \frac{1}{m_u+m_d}m_\pi^2 f_\pi,
\end{equation}
where $m_\pi$ is the pion mass and $V_{3d}$ is the spatial $3$-volume.
On the lattice, the matrix element is renormalized as
\begin{equation}
    \frac{1}{\sqrt{V_{3d}}}\sum_{\vec{x}}\langle \pi(m_\pi,\vec{0}) | P(0,\vec{x})|0\rangle =\frac{Z_S}{Z_P} \frac{1}{m_u+m_d}m_\pi^2 f_\pi,
\end{equation}
where $Z_S$ and $Z_P$ are the renormalization constants of scalar and pseudoscalar densities respectively. Since all of staggered, unimproved and improved KW actions are chiral, $Z_S/Z_P=1$.
The pion correlator on the lattice between time slices $x_0$ and $y_0$,  $G_{PP}(x_0,y_0)$, is defined as
\begin{equation}
    G_{PP}(x_0,y_0) \equiv \frac{1}{V_{3d}} \sum_{\vec{x}} \sum_{\vec{y}} \langle 0 | \bar{P}(x) P(y) | 0 \rangle \, \, .
\end{equation}
Defining $\tau \equiv x_0 - y_0$ , averaging over $y_0$, and using spectral decomposition, $\langle G_{PP}(y_0+\tau,y_0)\rangle_{y_0}$ acquires the following asymptotic form as $\tau \rightarrow \infty$:
\begin{align}
    & \langle G_{PP}(y_0+\tau,y_0)\rangle_{y_0}\\\notag
    &\xrightarrow{\tau\rightarrow\infty}\left|\frac{1}{\sqrt{V_{3d}}}\sum_{\vec{x}}\langle \pi(m_\pi,\vec{0}) | P(0,\vec{x}) | 0 \rangle\right|^2\frac{1}{2~m_\pi} e^{-m_\pi \tau}\\\notag
    =&\left|\frac{Z_S}{Z_P} \frac{1}{m_u+m_d} ~ m_\pi^2 ~ f_\pi\right|^2\frac{1}{2~m_\pi} ~ e^{-m_\pi \tau}\\\notag
    =&\frac{1}{2(m_u+m_d)^2} ~ m_\pi^3 ~ f_\pi^2 ~ e^{-m_\pi \tau}.
\end{align}
In our simulation, the correlator is obtained stochastically by
\begin{align}
    &C(\tau)\\\notag
    \equiv & C_0 \sum_{\vec{x},\vec{y},y'} \langle \gamma_5 D^{-1}[U](y,x) \gamma_5 D^{-1}[U](x,y')\eta(y')\eta^\dagger(y) \rangle_{y_0,\eta,U},
\end{align}
where $x \equiv (y_0+\tau, \vec{x})$ and the Gaussian noise vectors $\eta$ are restricted on time-slice $y_0$ (i.e. $y'=(y_0,\vec{y}')$) and satisfy 
\begin{equation}
\langle \eta_{\alpha,a}(x) \eta^*_{\alpha',a'}(x')\rangle_\eta = \delta_{\alpha,\alpha'}\delta_{a,a'}\delta_{xx'}
\end{equation}
with $\alpha,\alpha'$ and $a,a'$ being the spin and color indices respectively. In this work, sixteen $y_0$ timeslices are separated evenly across the temporal extent of the lattice and one noise vector per timeslice is used. The normalization constant $C_0$ is defined for staggered and KW actions respectively as
\begin{align}
    C_0^{\texttt{Staggered}}&\equiv \frac{1}{2}\frac{(m_u+m_d)^2}{V_{3d}}, \\
    C_0^{\texttt{KW}}&\equiv \frac{(m_u+m_d)^2}{V_{3d}}
\end{align}
so that  
\begin{equation}
C(\tau) \equiv 2(m_u+m_d)^2 \langle G_{PP}(y_0+\tau,y_0) \rangle_{y_0}.
\end{equation}
Hence $f_\pi$ and $m_\pi$ are readily obtained by the simple exponential form,
\begin{equation}
C(\tau) \xrightarrow{\tau \rightarrow \infty} m_\pi^3 ~f^2_\pi ~ e^{-m_\pi \tau}.
\end{equation}

\begin{figure} \centering 
    \includegraphics[width=.5\textwidth]{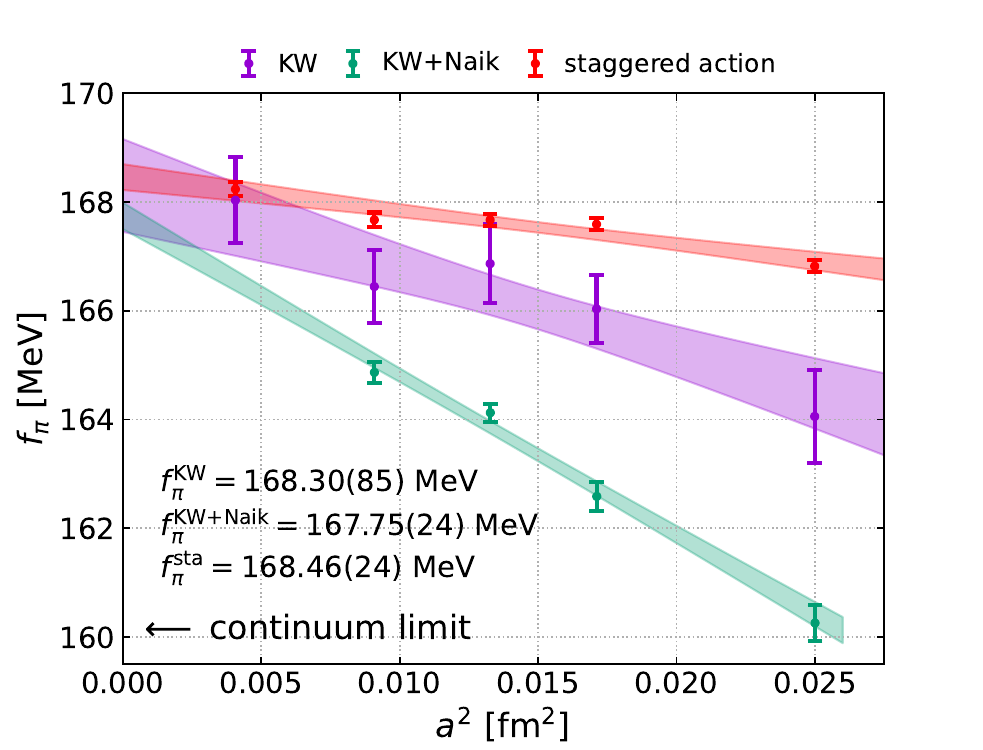}
    \caption{Continuum extrapolation of the pion decay constant $f_\pi$ for 
    the improved and unimproved KW action, as well as with the staggered 
    action.}
    \label{fig:pion_constant} 
\end{figure}

In Fig.~\ref{fig:pion_constant} we show the continuum extrapolation 
of $f_\pi$ measured using both the KW actions, as well as the 
staggered one. All measurements used the same 144 configurations. In all three cases the points shown are the results of a systematic analysis of the uncertainties,
where we varied the range of the fit, then combined the results with the histogram method \cite{Borsanyi:2020mff}. The same treatment was used to estimate the uncertainties in Fig.~\ref{fig:taste_splitting}.  We 
find that the continuum limit of the unimproved KW action, due to its 
larger uncertainty, agrees with both the Naik improved KW action and 
the staggered one. With our current results, a slight tension appears 
between the Naik improved and staggered continuum limits. We also 
note that, with the same statistics, the Naik improved action yields 
substantially less noisy results, compared to the unimproved one. 
Moreover, the continuum limit has a 3x smaller error than the 
unimproved one, even without the use of the finest lattice. This 
fact bodes well for the cost effectiveness of the Naik improvement in 
the KW action.

%  Mg5 , Mg0, Mg0gi, Mg5gi, M1
% Mg0^2 - Mg^2 = A 
% Mg0gi^2 - Mg0^2 =B
% Mg5gi^2 - Mg0gi^2 =B
% M1^2 - Mg5gi^2 =B
%  Mg5 , Mg0, Mg0gi, Mg0gi-Mg0+Mg0gi , 2*(Mg0gi-Mg0)+Mg0gi

% Mg0^2 - Mg5^2
% (Mg0^2 + Mc0g0^2 + Mc0g5^2)/3 - Mg5^2

%%%%%%%%%%%%%%%%%%%%%%%%%%%%%%%%%%%%%%%%%%%%%%%%%%%%%%%%%%%%%%%%%%%%%%%%%%%%%%%%
\section{\label{sec:taste}Taste structure}

We now wish to gauge the impact of taste breaking in both the 
unimproved and Naik improved KW actions. Thus, we investigate the 
mass-splittings of the ground states of various fermion channels that 
form parity partners in the spin-taste structure of the KW action 
\cite{Weber:2023kth}. 
For the na{\"i}ve fermion action, the $\gamma_5$ and $\gamma_0$ 
channels are mass-degenerate parity partners, but they are no longer 
degenerate at finite lattice spacing for the KW action 
\cite{Weber:2015oqf,Weber:2023kth}. Ref.~\cite{Weber:2023kth} 
identifies ``site-split'' fermion operators $c_0\gamma_5$ and 
$c_0\gamma_0$ as parity partners in the same spin structure as 
$\gamma_5$ and $\gamma_0$.

\begin{figure} \centering 
\includegraphics[trim=0 0 0 0, clip,width=.45\textwidth]{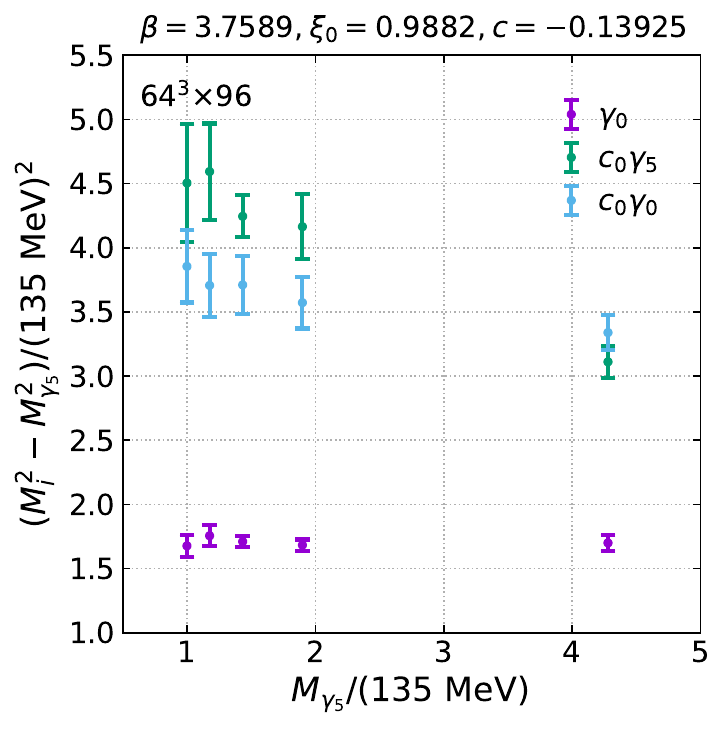}
\includegraphics[width=.45\textwidth]{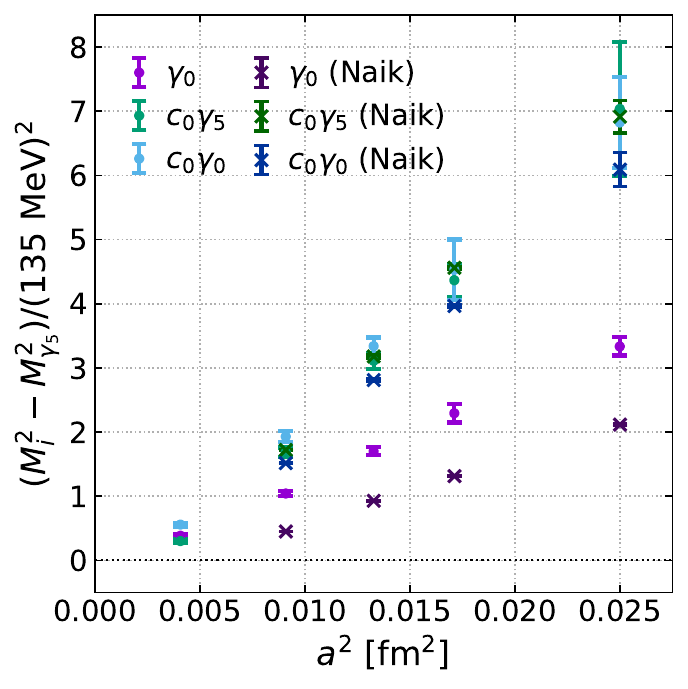}
\caption{(Top) $M_{\gamma_5}$ dependence of the taste-splitting 
$\Delta M_i^2 = M_i^2 - M_{\gamma_5}^2$ of three fermion channels in the 
unimproved KW action with $\beta=3.7589$. The $\gamma_0$ channel shows no 
$M_{\gamma_5}$ dependence even all the way down to the physical pion mass. 
The two site-split channels, $c_0\gamma_5$ and $c_0\gamma_0$, show a slight 
mass dependence, increasing as $M_{\gamma_5}$ decreases. (Bottom) Taste-
splitting of the same three fermion channels on both the unimproved and
Naik improved KW actions vs. the lattice spacing squared at fixed
$M_{\gamma_5} = 578.4 \ {\rm MeV}$. 
% The na{\"i}ve continuum limit for each taste appears to be negative.
\label{fig:taste_splitting} } 
\end{figure} 

We then consider the bare quark mass and lattice spacing dependence of the 
quadratic mass difference $\Delta M_i^2 = M_i^2 - M_{\gamma_5}^2$ for the 
three tastes $\gamma_0$, $c_0\gamma_5$, and $c_0\gamma_0$. 
We show in the top panel of Fig.~\ref{fig:taste_splitting} that 
$\Delta M_{\gamma_0}^2$ is very stable with respect to the 
pseudoscalar mass $M_{\gamma_5}$, even all the way down to the 
physical pion mass. This result agrees with the findings of 
Ref.~\cite{Weber:2015oqf}. For the $c_0\gamma_5$ and $c_0\gamma_0$ 
channels, the $M_{\gamma_5}$ dependence is small, although 
noticeable. The taste-splitting of both channels increases as 
$M_{\gamma_5}$ decreases.

In the bottom panel of Fig.~\ref{fig:taste_splitting} we show 
$\Delta M_i^2$ at tuned $c$ and $\xi_0$ with $M_{\gamma_5} = 578.4$ 
MeV as a function of the lattice spacing squared for all three 
tastes. Results for both the unimproved and Naik improved action are 
shown. The rightmost points in the top panel are the (unimproved) 
middle points in the bottom plot. We find that $\Delta M_i^2$ 
decreases for $a \rightarrow 0$ for all three tastes. We find the 
effect of the improvement to be present mostly for the $\gamma_0$ 
channel, while the site-split channels are virtually unchanged. The 
$\gamma_0$ channel mass-split is substantially reduced, from about 
$30\%$ on the coarsest lattice to more than $50\%$ on the finest 
simulated one. This is another argument in favour of the Naik 
improvement, which does seem to effectively reduce discretization 
errors and hence facilitate the continuum limit.

%%%%%%%%%%%%%%%%%%%%%%%%%%%%%%%%%%%%%%%%%%%%%%%%%%%%%%%%%%%%%%%%%%%%%%%%%%%%%%%%
\section{\label{sec:conclusions}Conclusions}

We have presented the results of a mixed-action study of the 
Karsten-Wilczek fermion action on 4stout staggered simulations, 
considering both an unimproved version, and one with a tree-level 
Naik improvement in the three spatial directions. Although tedious, a 
tuning procedure involving the fixing of two fermionic counterterms, 
plus the bare quark mass, was formulated, whereby the different 
parameters might be tuned almost independently. A hierarchy in the 
tuning of the parameters was also established, based on the relative 
ease of the procedure, noting that the $c$ parameter should be fixed 
first, then the anisotropy $\xi_0$, and finally the quark 
mass. Importantly, this means that most of the tuning can be carried 
out at much cheaper, larger values of the mass, before being
finalized at its physical value.

The investigation of the spatial Naik improvement showed that, with a
reasonable cost increase, much less noisy results are obtained. The 
cost increase of a single iteration of the solver, compared to the 
unimproved case, is roughly a factor 2 (for the spatial part only). 
Additionally, as shown in Fig.~\ref{fig:cost} as a function of the 
inverse pseudoscalar mass, the number of iterations needed is also 
increased, by about $20\%$. The rightmost point in the plots roughly 
corresponds to a physical pion mass. In light of these additional 
costs, Fig.~\ref{fig:pion_constant} can give us an idea of the 
comprehensive cost-effectiveness of the Naik improvement. Overall, 
with an approximate per-lattice cost increase slightly above 2, we 
obtain roughly 4x smaller errors on the individual lattices. In the 
continuum extrapolation, the error reduction is of about a factor 
3.5, even without simulating the finest lattice. Hence, we can 
conclude that the extra cost is more than compensated, and the Naik 
improved KW action should be used preferably.

\begin{figure} \centering 
    \includegraphics[width=.5\textwidth]{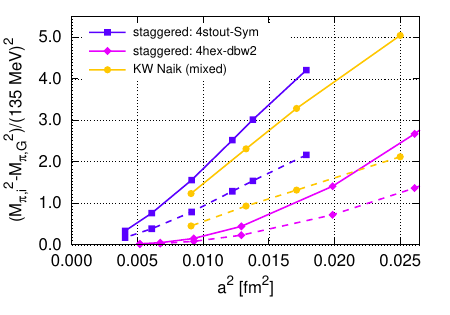}
    \caption{Taste breaking in the KW action, compared to our 4stout and 4HEX staggered actions. The solid lines indicate the normalized mean-square taste breaking, while the dashed lines correspond to the second-lowest lying states.}
    \label{fig:taste_breaking_world} 
\end{figure}

Additionally, as we showed in our discussion of the taste breaking 
features of the action, the improvement significantly reduces the 
mass-split of the $\gamma_0$ channel, although the same doesn't 
happen for the site-split channels. It is also worth considering how 
the KW action compares to currently used actions, in particular 
staggered ones. In Fig.~\ref{fig:taste_breaking_world} we show the 
taste breaking of our 4stout and 4hex staggered actions, compared to 
our findings for the improved KW action, as a function of the lattice 
spacing. First, it is to be noted that the KW action has only four 
tastes, compared to the sixteen of the staggered formulations. Shown 
in the plot are the squared-mass difference between a certain taste
and the Goldstone (G) one, normalized by the squared physical pion 
mass. We show for the KW case the $\gamma_0$ channel (dashed line) 
and the root-mean-squared mass-split (solid line). For the 
staggered actions, the dashed lines still indicate the second-
lowest lying state, and with solid lines the staggered tensor states. 
This is because these are the most numerous, and their mass is 
typically very close to the normalized-mean-square, for which they serve as good proxies. 

We find that both 
the normalized-mean-squared mass-split and the lowest lying channel 
in the improved KW sit slightly below the corresponding ones for the 
4stout action. We also recall that the gauge ensembles used in this mixed 
study were generated with the same 4stout action, so that the yellow and 
purple points only differ by the fermionic measurements. We leave the 
exploration of the use of a different gauge action, smearing and anisotropy
for future work.

In light of the results shown in this work, we can conclude that the 
Naik improved KW action is a valid alternative to currently employed 
staggered formulations, because of the lack of rooting (with 
$N_f=2$), as well as reasonable cost-effectiveness and a favourable 
taste-breaking structure.

\subsection*{Acknowledgments}
This work was partly funded from the DFG under the Project
No. 496127839.  This work is also supported by the
MKW NRW under the funding code NW21-024-A. 
The authors gratefully acknowledge the Gauss Centre for
Supercomputing e.V. (\url{www.gauss-centre.eu}) for funding
this project by providing computing time on the GCS
Supercomputer Juwels-Booster at Juelich Supercomputer
Centre and HAWK at HLRS Stuttgart.

\bibliography{thermo}

\end{document}

%% file: TableI.tex
\begin{table}
% \begin{tabular}{| p{0.25\linewidth} | p{0.25\linewidth}>{\centering} | p{0.05\textwidth}>{\centering} | p{0.05\textwidth}>{\centering} |}
\begin{tabular}{|c|c|c|c|c|}
\hline
$\qquad \beta \qquad$ & $\quad a$ [fm] $\quad$ & $\qquad c \qquad $ & $\qquad \xi_0 \qquad$ & $\quad a m_0 \quad$  \\
\hline
3.6736 & 0.1581 & -0.18797 & 0.9943 & 0.05074 \\ 
\hline
3.7089 & 0.1308 & -0.15733 & 0.9888 & 0.04131 \\ 
\hline
3.7589 & 0.1152 & -0.13925 & 0.9882 & 0.03590 \\ 
\hline
3.8360 & 0.0953 & -0.11703 & 0.9846 & 0.02937 \\ 
\hline
4.0126 & 0.0638 & -0.08417 & 0.9854 & 0.01896 \\ 
\hline 
\end{tabular} 
\caption{ 
\label{tab:tuned_values_noim}
The tuned values of $c$, $\xi_0$ and $m_0$ for the unimproved action on the 
lattices used in this work. The bare mass $m_0$ is tuned to obtain 
$M_{\gamma_5} = 578.4$ MeV.
}
\bigskip
% \begin{table}
% \begin{tabular}{| p{0.25\linewidth} | p{0.25\linewidth}>{\centering} | p{0.05\textwidth}>{\centering} | p{0.05\textwidth}>{\centering} |}
\begin{tabular}{|c|c|c|c|c|}
\hline
$\qquad \beta \qquad$ & $\quad a$ [fm] $\quad$ & $\qquad c \qquad $ & $\qquad \xi_0 \qquad$ & $\quad a m_0 \quad$ \\
\hline
3.6736 & 0.1581 & -0.09387 & 1.12846 & 0.06102 \\ 
\hline
3.7089 & 0.1308 & -0.07137 & 1.10280 & 0.04864 \\ 
\hline
3.7589 & 0.1152 & -0.06129 & 1.09496 & 0.04205 \\ 
\hline
3.8360 & 0.0953 & -0.04605 & 1.07397 & 0.03373 \\ 
\hline 
\end{tabular} 
\caption{ 
\label{tab:tuned_values_impr}
The tuned values of $c$, $\xi_0$ and $m_0$ for the Naik improved action on 
the lattices used in this work. The bare mass $m_0$ is tuned to obtain 
$M_{\gamma_5} = 578.4$ MeV.
}
\end{table}